\documentstyle[11pt,cite]{article}
\newcommand{\mr}[1]{{\mathrm {#1}}}

\newcommand{\alse}{\alpha_s\;}

\newcommand{\epem}{e$^{+}${\rm e}$^{-}\;$}

\newcommand{\bare}[1]{\overline{#1}}
\newcommand{\reff}[1]{(\ref{#1})}
\newcommand{\be}{\begin{equation}}
\newcommand{\ee}{\end{equation}}
\newcommand{\bc}{\begin{center}}
\newcommand{\ec}{\end{center}}
\newcommand{\bd}{\begin{description}}
\newcommand{\ed}{\end{description}}
\newcommand{\bit}{\begin{itemize}}
\newcommand{\eit}{\end{itemize}}
\newcommand{\ben}{\begin{enumerate}}
\newcommand{\een}{\end{enumerate}}
\begin{document}
\begin{flushright}
PRA-HEP/95-05
\end{flushright}
\begin{center}
\vspace*{0.5cm}
\noindent
{\huge \bf On dynamically generated parton distribution functions and
their properties}\\
\vspace*{0.4cm}
{\large Ji\v{r}\'{\i} Ch\'{y}la} \\
\vspace*{0.4cm}
Institute of Physics, Academy of Sciences of the Czech Republic \\

\vspace*{1cm}
Abstract \\
\end{center}
\noindent
The idea of ``dynamically'' generated parton distribution functions,
based on regular initial conditions at low momentum scale, is
reanalyzed with particular emphasize paid to its compatibility with the
factorization mechanism. Basic consequences of this approach are
discussed and compared to those of the conventional approach, employing
singular initial distribution functions.

\section{Introduction}
One of the highlights \cite{Eisele,Mueller} of the recent International
Conference on High Energy Physics in Brussels has been the remarkable
success of ``dynamically'' generated parton distribution functions
(DGPD)
advocated by Gl\"{u}ck, Reya and Vogt (GRV)
\cite{GRV1,GRV2,GRV3,GRV4,GRV5,GRV6}, in predicting the rapid rise of
proton structure function $F_2^{\mr{ep}}(x,Q)$ at low $x$, observed at
HERA \cite{H1,ZEUS}.
\footnote{Throughout the paper the term ``distribution'' stands
for distribution function.},

The GRV group is one of three main groups
(the other two being the Durham (MRS)
\cite{MRS1,MRS2,MRS3} and CTEQ \cite{CTEQ1,CTEQ2} ones), which
systematically analyze hard scattering data within the framework of
perturbative QCD. What distinguishes GRV approach from those of the
other two groups is their claim that the DGPD are more than just
parameterizations of our inability to compute structure functions
directly from first principles. GRV argue that by imposing certain
condition on the initial parton distributions at low momentum scale, one
obtains more predictive results. Without this
additional theoretical input the conventional parameterizations, using a
moderate initial scale $Q_0\approx 2$ GeV, are unstable when
extrapolated to low $x$ region. For that reason both the MRS
and CTEQ groups usually present several sets of such parameterizations,
differing just in low $x$ region.

The idea of DGPD is intuitively
appealing and actually almost as old as QCD itself \cite{GRV1}.
Confronted with growing amount and variety of data, it has, however,
undergone significant modifications \cite{GRV2,GRV3,GRV5} and in the
process lost most of its original appeal.  As the GRV approach
relies on very low initial scale in the range $0.5-0.6$ GeV, it has been
met with reservations and scepticism  \cite{Forshaw,Durham1,Durham2}.
In response to this criticism and in order to bring further arguments in
favour of their approach, GRV have included in
their recent paper \cite{GRV5} an extensive discussion of several of
these points.

To relate physics of short distances, the true realm of perturbative
QCD, to that of distances comparable to the proton size would
certainly represent a major achievement. The purpose of this paper is
to discuss whether this can really be done in the way suggested in
\cite{GRV1,GRV2,GRV3,GRV4,GRV5,GRV6}.
Throughout this paper I shall concentrate on the analysis
of the basic assumptions and consequences of the GRV approach,
with only occasional reference to comparison with experimental data.

The paper is organized as follows. In the next section I shall briefly
recall the development of the idea of DGPD, from its inception
\cite{GRV1} up to the present status \cite{GRV5}. In Section 3 the
applicability of perturbative QCD at distances as large as 0.4 fm will
be discussed. In particular I shall comment on the implications and
interpretation of recent lattice calculations \cite{Luscher}, quoted in
\cite{GRV5}. The indispensable role of power corrections in going from
short distances (where partons live) to distances comparable to the
proton size (where the appropriate degrees of freedom are the
constituent quarks) is emphasized in Section 4.
In Section 5 the compatibility of the DGPD with
the factorization mechanism is discussed in detail. In particular it is
shown why it is very difficult for gluons and sea quarks to be
valence--like. This discussion also shows how the conventional
parameterizations, based on singular input distributions, avoid this
problem. In Section 6 results of the conventional approach in the small
$x$ region are briefly reviewed and cast into a simple form suitable for
the comparison with the DGPD. This comparison, carried out in
Section 7, identifies two basic signatures of the DGPD.
Throughout the paper I adopt the notation in which the QCD couplant
$a=\frac{\alse}{\pi}$ satisfies the usual RG equation
\be
\frac{\mr{d}a(M,\mr{RS})}{\mr{d}\ln M}=
-ba^2(M,\mr{RS})\left(1+ca(M,\mr{RS})
+c_2a^2(M,\mr{RS})+\cdots\right),
\label{RG}
\ee
where $b,c$ are the first two, universal, coefficients
\footnote{Assuming QCD with 3 colors and $n_f$ massless quark flavors.}
\be
b= \frac{33-2n_f}{6},\;\;\;c= \frac{153-19n_f}{66-4n_f},
\label{bc}
\ee
while all the higher order coefficients $c_k;k\ge 2$ in \reff{RG} are
free parameters, defining the so called renormalization convention
(RC) \cite{PMS}.  Together with the specification of the initial
condition on the solution of \reff{RG} they define the renormalization
scheme (RS).

\section{The evolution of the idea of dynamically generated partons}
The original idea of \cite{GRV1} was to generate
parton distributions at large momentum scales, where experimental data
are available, by means of the DGLAP leading order
\footnote{In later  GRV papers also the NLO DGLAP evolution equations
were used.},
leading twist evolution equations, starting at some small momentum
scale \footnote{In the rest of this paper $Q_0$ is used for the general
initial scale, while the symbol $\mu$ is reserved for the initial scale
within the GRV approach, i.e. the one at which the parton distributions
become valence--like.}
$\mu\approx 0.55$ GeV from purely valence--like quark distributions,
with vanishing light sea and gluon ones
\footnote{In GRV approach heavy quarks $c,b$ and $t$ are not considered
as intrinsic partons in the nucleon, but are produced from intrinsic
gluons via the boson--gluon fusion mechanism \cite{GRV5}.}
\be
G(x,\mu)=\bare{u}(x,\mu)=\bare{d}(x,\mu)=\bare{s}(x,\mu)=s(x,\mu)=0.
\label{initial}
\ee
The quark distributions at the initial scale $\mu$, obtained by backward
evolution from measured structure function $F_2^{eN}(x,Q)$ at $Q^2_0=3$
GeV$^2$, were constrained to satisfy quark number sum rule
\be
\int^1_0 \mr{d}x
\left[u(x,\mu)+d(x,\mu)\right]=3,
\label{flavorsumrule}
\ee
which provides a fundamental bridge between the parton model of Feynman
and the old nonrelativistic ``quasinuclear colored model'' of
Gell--Mann, Zweig, Greenberg, Lipkin and others. The scale $\mu$ was
fixed by imposing the momentum sum rule
\be
\int^1_0 \mr{d}x
x\left[u(x,\mu)+d(x,\mu)\right]=1.
\label{sumrule}
\ee

In 1976 there were too few data on hard scattering processes to test
the DGPD thoroughly. With more and increasingly accurate
data becoming available in late eighties, the GRV were forced to modify
their original idea by allowing nonvanishing valence--like gluon
distribution $G(x,\mu)$ as well \cite{GRV2}. Moreover, $G(x,\mu)$ was
related to the input valence--like quark distributions as follows:
\be
G(x,\mu) = \frac{n_G}{3}\left[u(x,\mu)+d(x,\mu)\right],\;\;\;\;
\bare{q}(x,\mu)=0.
\label{qbar}.
\ee
In \cite{GRV2} $\mu$ and $n_G$ were fixed by means of the momentum sum
rule at the initial scale $\mu$, now including also the gluon
contribution, together with the comparison of theoretical predictions
with data on direct photon production. They got about the same $\mu$ as
in \cite{GRV1} and $n_G=2$.

Confronted with still more data GRV had finally to
include in their initial parton distributions also the valence--like
nonstrange sea \cite{GRV3} so that the momentum sum rule now reads
\be
\int^1_0 \mr{d}x x\left[u_v(x,\mu)+d_v(x,\mu)+2\bare{u}(x,\mu)+
2\bare{d}(x,\mu)+G(x,\mu)\right]=1.
\label{sumruleall}
\ee
In one of their latest NLO global analysis, published in \cite{GRV5},
$\mu=0.58$ GeV and
\begin{eqnarray}
xu_v(x,\mu) & = & 0.988x^{0.543} \left(1+1.58\sqrt{x}+2.58
x+18.1x^{3/2}\right)(1-x)^{3.38}, \label{uv}  \\
xd_v(x,\mu) & = &
0.182x^{0.316}\left(1+2.51\sqrt{x}+25.0x+11.4x^{3/2}\right)(1-x)^{4.113},
\label{dv} \\
x(\bare{u}+\bare{d})(x,\mu) & = & 1.09x^{0.3}(1+2.65x)(1-x)^{8.33},
 \label{ubarplusdbar} \\
xG(x,\mu) & = & 26.2x^{1.9}(1-x)^{4.0}, \label{glue}   \\
xs(x,\mu) & = & x\bare{s}(x,\mu)=0.\label{ss}
\end{eqnarray}

\section{Does perturbative QCD make sense at 0.4 fermi?}
The initial scale $\mu\doteq 0.58$ GeV corresponds to a distance $0.37$
fermi. As pointed out by a number of authors
\cite{Forshaw,Durham1,Durham2} such distances are probably too large for
a meaningful purely perturbative treatment. It is fair to say that GRV
do not trust their results at such low momentum scales but claim they
are good approximations only above somewhat higher scale
$\mu_{\mr{pert}}\doteq 0.75$ GeV \cite{GRV5}. However, even the latter
value seems too low for the applicability of leading twist, low order
(LO or NLO) perturbative QCD. The following paragraphs are intended to
throw some light on this problem.

In this context let me first comment on the results of recent lattice
calculation \cite{Luscher}, quoted in \cite{GRV5}.
According to \cite{GRV5} these results ``confirm the perturbative
NLO (2 loop) predictions for $\alpha_s(Q)$ down to $Q=0.55$ GeV.'' This
claim relies on Fig. 1, taken from \cite{Luscher}, where the results of
nonperturbative lattice evaluation
\footnote{The definition of $\alpha_s^{\mr{latt}}$ used in
\cite{Luscher} is rather involved and is therefore not mentioned here.}
of the QCD running coupling $\alpha_s^{\mr{latt}}(q)$ is plotted as a
function of $q$ in the region $q\in (0.5,14)$ GeV. The agreement
between $\alpha_s^{\mr{latt}}(q)$ and the curve corresponding to 2--loop
perturbative $\beta$--function, i.e. the solution of \reff{RG} with only
the first two universal terms on its r.h.s., down to 0.5 GeV is indeed
remarkable. However, what is demonstrated by Fig. 1 is merely the fact
that a particularly defined lattice couplant
$\alpha_s^{\mr{latt}}$ coincides
with the couplant defined in the so called 't Hooft RC \footnote{In this
RC all nonunique $\beta$--function coefficients $c_j,j\ge 2$ are set to
zero by definition.} down to $q=0.55$ GeV. It is worth emphasizing
that even on the lattice there is no unique ``nonperturbative''
$\beta$--function and, consequently, no unique nonperturbative couplant
$\alpha_s^{\mr{latt}}$!  While asymptotic freedom of QCD guarantees that
at short distances couplants in different RS coincide, they may be
arbitrarily far apart at large ones. Fig. 1 contains an interesting
evidence for the closeness of two (out of an infinite number)
definitions of the couplant, but tells us nothing about the
applicability of perturbation theory in any of them. The authors of
\cite{Luscher} are well aware of this limitation and on page 495 of
\cite{Luscher} therefore write:  ``We would like to emphasize, however,
that our results do not prove that perturbation theory provides a good
approximation to all quantities of interest up to couplings as large as
3.48. Such a general statement is bound to be false and the running
coupling in our scheme may very well turn to be an exceptional case.''
\footnote{The value 3.48 quoted above corresponds to $g^2$.}

To illustrate the importance of nonperturbative effects at distances
$0.3-0.4$ fermi, let us consider the magnitude  $F(r)$ of the force
between two static quarks at a distance
$r$. This quantity has been extensively studied on the lattice and is
usually written as the sum
\be F(r)=\frac{4}{3}\frac{\alpha_s^{q\bare{q}}(r)}{r^2}+\kappa=
F_p(r)+F_{np},
\label{force}
\ee
where the first term, dominant at short distances, comes purely from
perturbation theory while the second describes the nonperturbative, long
range confining force with $\kappa$ denoting the string
tension. The couplant $\alpha_s^{q\bare{q}}$ in the numerator of
\reff{force} is related to $\alpha_s^{\mr{latt}}$, mentioned above, as
follows \cite{Luscher}
\be
\alpha_s^{q\bare{q}}(M)=\alpha_s^{\mr{latt}}(M)\left(1+k_1
\alpha_s^{\mr{latt}}(M) +\cdots\right),\;\;\; k_1=1.33776.
\label{relation2}
\ee
Evaluating both terms in \reff{force} for
$\kappa=(0.48\;\mr{GeV})^2$ and
two values of $M$ ($M=\mu=0.55$ GeV, corresponding to $r=0.37$ fm and
$M=\mu_{\mr{pert}}=0.75$ GeV, corresponding to $r_{\mr{pert}}=0.27$ fm),
using the values of $\alpha_s^{\mr{latt}}$ from \cite{Luscher}, we find
\be
F_p(0.37\;\mr{fm})\doteq 0.17\;\mr{GeV}^2,\;\;
F_{p}(0.27\;\mr{fm})\doteq 0.27\;\mr{GeV}^2,\;\;
F_{np}\doteq 0.24\;\mr{GeV}^2.
\label{F75}
\ee
This is the kind of comparison which really tells us how important
is the perturbative contribution to a particular physical quantity,
in this case the interquark force \reff{force}. For this quantity
the nonperturbative contribution clearly
dominates over the perturbative one at the distance $r=0.37$
fm and is roughly equal to it at $r=0.27$ fm, the distance at which
dynamical perturbative predictions should, according to
\cite{GRV5}, ``become reliable and experimentally relevant''. The above
example is merely an illustration and the relative importance of
perturbative and nonperturbative parts may well depend on the physical
quantity in question, but it at least gives some indication that at
$\mu_{\mr{pert}}\doteq 0.75$ GeV perturbation contributions can
hardly be expected to be a good approximation to the full results.

One of the basic features of perturbation theory
at low momentum scales (large distances) is the increasing sensitivity
of finite order perturbative approximants to the choice of the
renormalization and factorization schemes and renormalization and
factorization scales
\footnote{In this note these two in principle
different scales will be identified and the dependence on the choice of
factorization scheme disregarded.},
which makes perturbative predictions
progressively more ambiguous in this region.
To illustrate the crucial importance at large distances of higher order
terms in purely perturbative expansions let me briefly recall the
essence of Ref. \cite{Paul}.
There the familiar $R$-ratio in \epem annihilation into hadrons at the
center of mass energy $Q$
\be
R_{\mr{e}^+\mr{e}^{-}}(Q) \equiv
\frac{\sigma(\mr{e}^+\mr{e}^-\rightarrow \mr{hadrons})}
{\sigma(\mr{e}^+\mr{e}^-\rightarrow \mu^+\mu^-)}=
3\left(\sum_{i=1}^{n_f} e_i^2\right)(1+r(Q)),
\label{epem}
\ee
where in perturbative QCD
\be
r(Q)= a(M,\mr{RS})\left[1+r_{1}(Q/M,\mr{RS})a(M,\mr{RS})+
 r_{2}(Q/M,\mr{RS})a^2(M,\mr{RS})+\cdots\right]
\label{r(Q)}
\ee
is investigated in the infrared region.
Their analysis has two important ingredients
\bit
\item the use of the NNLO approximation to \reff{r(Q)}, with $c_2$
chosen by means of the PMS \cite{PMS},
\item smearing of $R_{\mr{e}^+\mr{e}^{-}}(Q)$ over some interval
$\Delta$ of $Q$:
\be
\bare{R}_{\mr{e}^+\mr{e}^{-}}(Q,\Delta)\equiv
\frac{\Delta}{\pi}\int_0^{\infty}\mr{d}s\frac{
R_{\mr{e}^+\mr{e}^{-}}(\sqrt{s})}{(s-Q^2)^2+\Delta^2} \label{smearing}
\ee
\eit
The second ingredient of this
procedure is vital as the detailed structure of
$R_{\mr{e}^+\mr{e}^-}(Q)$ in the resonance regions is clearly beyond
the reach of perturbative QCD. Nevertheless the fact that after the
smearing and for not too small values of $\Delta$, the agreement between
\reff{smearing} and the data is quite good, as documented by Fig. 6 of
\cite{Paul}, is remarkable. This agreement depends crucially on the
fact that $c_2$, as chosen by the PMS, is negative, since for positive
$c_2$ the NNLO perturbative expansion for \reff{smearing} blows up in
the IR region, as does the NLO one. Also the magnitude of $c_2$ is
important as it determines the magnitude of $\bare{R}$ in the IR region.
The success of such a procedure might look suspicious as
in QCD sum rule approach resonance parameters (and thus
also their contribution to the smeared spectra), are dual to power
corrections. The observed agreement between data and NNLO perturbative
approximation in the PMS approach can be interpreted as a signal that by
an appropriate choice of $c_2$ (or in general of the RC), it may be
possible to include in some sense also the effects of these power
corrections. Such an interplay of perturbative and nonperturbative
contributions is quite plausible, as they actually coexist
within the OPE.  I have mentioned the analysis of \cite{Paul} merely to
emphasize that at large distances the inclusion of NNLO perturbative
terms is probably indispensable for meaningful and reasonably complete
description of physical quantities.

The message of the previous paragraph may also have some relevance for
``perturbative stability'' observed within the GRV approach \cite{GRV5}
in the comparisons of the LO and NLO approximations to the leading twist
DGLAP equations. Since the LO approximation cannot be associated
with any well--defined RS, and the importance of higher order terms
depends sensitively on the RS, such a comparison makes little
quantitative sense. Only by comparing the NLO and the NNLO can such an
information be obtained. Unfortunately, there are only a few simple
quantities for which the NNLO calculations are available. One of them is
the perturbative part \reff{r(Q)} of the ratio \reff{epem}. Evaluating
just for illustration the first three known terms of \reff{r(Q)} at
$\mu_{\mr{pert}}\doteq 0.75$ GeV, for three flavors and in the
conventional $\bare{\mr{MS}}$ RS, we find that they are roughly in the
ratio $1:0.22:0.33$! Not only is there no sign of perturbative
stability for this quantity, but at such low scales the complicated
problem of the presumable divergence of perturbation expansions in fixed
RS becomes of utmost phenomenological importance.

\section{The interpretation of input parton distributions}
Let us now turn to the interpretation of parton distributions at the
initial scale scale $\mu$. While in \cite{GRV1} they were considered to
correspond to three {\em constituent} quarks, according to the  latest
paper \cite{GRV5} the initial valence--like quark and gluon
distributions ``should rather be identified with {\em current} quark
content of hadrons.'' This slight but crucial shift of interpretation
should justify why the sea and gluon distributions do not vanish at the
initial scale $\mu$, as would seem appropriate for real constituents of
the proton. Such an extension would still be reasonable were the
additional sea quark and gluon valence--like initial distributions
small admixtures to the basically three valence quark component of the
nucleon.

In \cite{GRV2} the fact that two valence
``constituent'' gluons at $\mu\approx 0.5$ GeV were required by the data
was considered ``fine'' as ``they may combine to give color
and spin singlets as is required for the nucleon.''. However,
the success of the conventional SU(6) quark model relies on the fact
that \underline{all} color singlet combinations of three constituent
quarks do exist in the nature, not only some of them! But in the color
singlet state
of three quarks and two gluons the latter do \underline{not} have to
couple to a color singlet. The system of three quarks and
two gluons would have  much richer spectrum of low--lying states than
the state of mere three quarks. It would be a kind of ``hybrid'' states,
suggested in the early eighties, but never found. In order to avoid
these problems arguments would have to be invented to show why
these two constituent gluons must (and not only may) couple
to a color and spin singlet. I am not aware of any such argument.

Similar problems arise for the initial distributions in \cite{GRV5},
summarized at the end of Section 2. Integrating over the initial
distributions without the prefactor $x$ to get the probabilities, we
find
\footnote{There is an misleading claim in \cite{GRV5}
that their input distribution functions imply that ``proton consists
dominantly of valence quarks and valence--like gluons, with only 10\%
q$\bare{\mr{q}}$ excitations (sea quarks).'' The mentioned 10\% concerns
the momentum fraction carried by sea quarks and antiquarks,
not the probabilities themselves.}
\be
\int^1_0 G(x,\mu)\mr{d}x \doteq 1,\;\;\;
\int^1_0 \left(\bare{u}(x,\mu)+\bare{d}(x,\mu)\right)\mr{d}x
\doteq 1.6.
\label{integraly}
\ee
The more accurate and copious data used in \cite{GRV5} thus lead to the
result that the initial parton distributions describe a system composed
of 2.72 $u$ quarks, 1.88 $d$ quarks, 0.72 $\bare{\mr{u}}$ antiquark,
0.88 $\bare{\mr{d}}$ antiquark and about one valence gluon. So again
valence sea (anti)quarks and gluons are by no means a small admixture
but on the contrary provide a dominant component of the initial parton
distributions.

According to GRV this is no cause for concern as perturbation theory is
not expected to hold at the initial scale $\mu\approx 0.55$ GeV, but
only above $\mu_{\mr{pert}}\doteq 0.75$ GeV. If, however, initial parton
distributions do not describe at least approximately physics at the
scale $\mu$, what justifies then the adjective ``dynamical''? In
particular why to impose the fundamental sum rules \reff{flavorsumrule}
and \reff{sumrule}, or \reff{sumruleall},
upon which the GRV approach is based? For instance, if the initial
parton distributions $q(x,\mu),\bare{q}(x,\mu)$ and $G(x,\mu)$ are
irrelevant for physics, why should they satisfy the momentum sum rule
\reff{sumruleall}, which expresses the fact that quarks, antiquarks and
gluons carry together the whole momentum of the proton? And why should
there be just two $u$ and one $d$ quarks at the scale $\mu$? The answer
GRV offer exploits the invariance of these sum rules under the LO
DGLAP evolution equations and the fact their they do hold at short
distances. This reasoning goes, however, against the very spirit of
the DGPD, which I see in the possibility to use some \underline{known}
features of physics at long distances in order to \underline{predict}
physics at short ones. Imposing restrictions on the initial
distributions in the situation when the latter have no physical
meaning is basically a mathematical game with little physical content.
Nevertheless, such a game can have nontrivial consequences, which,
if confirmed by data, would signal some interesting physics behind
the GRV approach and would justify it a posteriori.

In my view the relation between constituent quarks and partons cannot
be described by leading twist DGLAP evolution equations, even if these
were taken to all orders.
\footnote{For related discussion see \cite{MRS4}.}
As pointed out in Section 2, the distance
$\approx 0.4$ fm, which corresponds to $\mu\approx 0.55$, is not much
smaller that the approximate size of three constituent quark in
the proton. The whole point of introducing the concept of constituent
quark is that it represents the {\em effective} degree of freedom
appropriate for describing the proton at distances comparable to its
size. Contrary to the current quarks, which are associated
with quark fields entering directly the QCD lagrangian, constituent
quarks have no such firm basis and are merely an intuitively introduced
concept, a kind of \underline{quasiparticle}, which is reasonably
well--defined only in the nonrelativistic quark model! The
phenomenological success of this model suggests that in low momentum
transfer processes proton behaves approximately as composed of three
constituent quarks, each with a mass of about 300 MeV. Constructing the
fields corresponding to constituent quarks from those of the current
quarks could probably be compared to the Bogolubov
transformation between electrons and fermionic quasiparticles in the BCS
theory of superconductivity. We expect the transition from short
distances, where current partons are the right degrees of freedom, to
large ones, where constituent quarks are the appropriate effective
degrees of freedom, to be smooth, but involve complicated multiparton
effects. In the framework of OPE such effects are described by
multiparton distributions, which naturally appear as part of power
corrections \cite{Ellis}. We cannot hope to get constituent quarks at
low scale from quarks, antiquarks and gluons at large scales merely by
means of the leading twist DGLAP evolution equations.

\section{Valence--like initial distributions and factorization}
In this Section the compatibility of the GRV approach with the mechanism
of factorization of parallel singularities will be discussed. In order
to make the discussion as clear as possible and to concentrate on the
essence of the problem, a number of simplifications will be made.

First, we shall be primarily interested in the low $x$ domain,
roughly $x\le 10^{-2}$. Secondly, all the considerations will be done
within the LO DGLAP equations. The inclusion of the NLO corrections is
not essential for any of the points discussed below. In the LO
approximation the basic quantity of interest, proton structure
function $F_2^{\mr{ep}}(x,Q^2)$, can be expressed in term of elementary
quark distributions as follows:
\be
F_2^{\mr{ep}}(x,Q^2) = x
\left[
\frac{4}{9}\left(u(x,Q^2)+\bare{u}(x,Q^2)\right)+
\frac{1}{9}
\left(d(x,Q^2)+\bare{d}(x,Q^2)+s(x,Q^2)+\bare{s}(x,Q^2)\right)
\right].
\label{F2}
\ee
Assuming SU(3) symmetry of the proton
sea the r.h.s. of \reff{F2} can be written as
a combination of the valence $u_v$, $d_v$ and the common sea
$D\equiv u_{\mr{sea}}=d_{\mr{sea}}=s_{\mr{sea}}$
distributions
\be
F_2^{\mr{ep}}(x,Q^2) =
\frac{4}{9}xu_v(x,Q^2)+\frac{1}{9}xd_v(x,Q^2)+
\frac{4}{3}xD(x,Q^2),
\label{F2SU3}
\ee
or, alternatively, as a sum of separate contributions from $u$ and $d$
quarks
\be F_2^{\mr{ep}}(x,Q^2) =
\left(\frac{4}{9}xu_v(x,Q^2)+\frac{4}{3}xD^{(u)}(x,Q^2)\right)+
\left(\frac{1}{9}xd_v(x,Q^2)+\frac{4}{3}xD^{(d)}(x,Q^2)\right).
\label{F2ep}
\ee
In terms of conventional
moments of various functions (distribution, branching, etc.)
\be
f(n,Q)\equiv \int^1_0 x^{n} f(x,Q)\mr{d}x
\label{moment}
\ee
the LO DGLAP evolution equation for the nonsinglet quark distribution
reads
\be
\frac{\mr{d}q_{\mr{NS}}(n,Q)}{\mr{d}\ln Q}=\frac{\alpha_s(Q)}{\pi}
P^{(0)}_{qq}(n)q_{\mr{NS}}(n,Q),
\label{NS}
\ee
while for the quark singlet and gluon distributions we have a
system of coupled equations
\begin{eqnarray}
\frac{\mr{d}G(n,Q)}{\mr{d}\ln Q} & = & \frac{\alpha_s(Q)}{\pi}
\left[P^{(0)}_{GG}(n)G(n,Q)+P^{(0)}_{Gq}(n)(q(n,Q)+\bare{q}(n,Q))
\right],  \label{GGG} \\
\frac{\mr{d}(q(n,Q)+\bare{q}(n,Q))}
{\mr{d}\ln Q} & = & \frac{\alpha_s(Q)}{\pi}
\left[2n_f P^{(0)}_{qG}(n)G(n,Q)+
P^{(0)}_{qq}(n)(q(n,Q)+\bare{q}(n,Q))\right].
\label{qq}
\end{eqnarray}
As in the small $x$ region the evolution of the gluon distribution is
driven by the branching $G\rightarrow G+G$, we shall furthermore
drop the second term on the r.h.s. of \reff{GGG}. Moments of the gluon
distribution satisfy then the same kind of differential equation as
quark nonsinglet distribution:
\be
\frac{\mr{d}G(n,Q)}{\mr{d}\ln Q}=\frac{\alpha_s(Q)}{\pi}
P^{(0)}_{GG}(n)G(n,Q)
\label{appG}
\ee
and therefore also the corresponding solutions have the same form
\begin{eqnarray}
q_{\mr{NS}}(n,Q) & = &
A_{\mr{NS}}(n)\left[\frac{ca(Q)}{1+ca(Q)}\right]
^{-P^{(0)}_{qq}(n)/b}, \label{ANS} \\
G(n,Q) & = &
A_{G}(n)\left[\frac{ca(Q)}{1+ca(Q)}\right]
^{-P^{(0)}_{GG}(n)/b}, \label{AG}
\end{eqnarray}
where $A_{\mr{NS}}(n),A_{G}(n)$ are
unique finite constants, determining the
asymptotic behaviour of the moments $q_{\mr{NS}}(n,Q), G(n,Q)$ as
$Q\rightarrow \infty$.  According to the factorization mechanism
\cite{Politzer1} these constants contain all the information on long
range properties of the nucleon, uncalculable in perturbation theory.
They represent one way of specifying the boundary conditions on the
solution of evolution equations \reff{NS} and \reff{appG}.
Another and almost, but not entirely, equivalent way
follows from taking the ratio of \reff{ANS} and \reff{AG} at two
different scales. In this case boundary conditions are specified at
finite initial $Q_0$ and we have
\begin{eqnarray}
q_{\mr{NS}}(n,Q) & = & q_{\mr{NS}}(n,Q_0)
\left[\frac{a(Q_0)}{a(Q)}\frac{1+ca(Q)}{1+ca(Q_0)}\right]^
{P_{qq}^{(0)}(n)/b},
\label{Q0QNS} \\
G(n,Q) & = & G(n,Q_0)
\left[\frac{a(Q_0)}{a(Q)}\frac{1+ca(Q)}{1+ca(Q_0)}\right]^
{P_{GG}^{(0)}(n)/b}.
\label{Q0QG}
\end{eqnarray}
Let me first discuss the compatibility of the initial valence--like
distributions with the
factorization on the simpler case of the gluon density $G(x,Q)$. Note
that although they are of the same form, there is a profound difference
between the solutions \reff{ANS} and \reff{AG} for $n=0$ (the moment
giving the integral over the parton distributions). This is due to the
fact that while $P^{(0)}_{qq}(0)=0$, $P_{GG}^{(0)}(0)=+\infty$!
The former is a consequence of quark number conservation in the
$q\rightarrow q+G$ branching, while the latter comes from the
$\frac{1}{x}$ spectrum of soft gluons. Eq. \reff{ANS} implies that the
integral over the nonsinglet quark density,
$q_{\mr{NS}}(0,Q)=A_{\mr{NS}}$ is independent of $Q$ and provided it is
finite at some $Q$ it stays so everywhere. For the gluons there are two
possibilities:
\bd
\item {\bf a)}
$A_G(0)>0$ and then $G(x,Q)$ \underline{cannot} be
valence--like at any $Q$ for which $a>0$ since
\be
G(0,Q)=A_{G}(0)\left[\frac{1+ca(Q)}{ca(Q)}\right]^{+\infty}=
A_{G}(0)(+\infty)=\infty,
\label{infty}
\ee
\item {\bf b)}
$A_G(0)=0$, in which case \reff{infty} is ill--defined, but
\reff{Q0QG} applied to $n=0$ can still be used
\be
G(0,Q)= G(0,Q_0)
\left[\frac{a(Q_0)}{a(Q)}\frac{1+ca(Q)}{1+ca(Q_0)}\right]^{+\infty}.
\label{appliedto0}
\ee
If we now impose valence--like behavior on the gluon distribution at the
initial $Q_0=\mu$, i.e. assume finite $G(0,\mu)$, \reff{appliedto0}
implies, due to monotonous behavior of the square bracket
as a function of $Q$, that $G(0,Q)$ is a \underline{discontinuous}
function of the factorization scale $Q$ at $Q=\mu$:
\begin{eqnarray}
G(0,Q) & = & +\infty, \;\;\forall Q>\mu  \label{Q>Q0} \nonumber \\
G(0,\mu) & = & \mr{const.}>0  \label{Q0} \\
G(0,Q) & = & 0\;\;\;\;\;\;\;\; \forall Q<\mu. \label{Q<Q0} \nonumber
\end{eqnarray}
Similar discontinuity at the initial scale
$\mu$ appears also for sea quarks and antiquarks.
\ed
In the case b), realized in the GRV approach, it is difficult to
understand why the divergence of $P^{(0)}_{GG}(0)$, which is a purely
perturbative phenomenon, should be accompanied by the vanishing of
nonperturbative quantity $A_{G}(0)$. For instance, in $4-\epsilon$
dimensions $P^{(0)}_{GG}(n,\epsilon)$ is finite and there is no
obvious reason to expect $A_{G}(0,\epsilon)=0$. Sending
$\epsilon\rightarrow 0$ we understand why
$P^{(0)}_{GG}(0,\epsilon)\rightarrow \infty$, but why should
simultaneously $A_{G}(0,\epsilon)\rightarrow 0$? Although the lowest
twist contribution may not provide a reliable description of the proton
at the initial scale $\mu$, we are not free to impose arbitrary
constraints on its properties.
Despite these reservations, the case b) is certainly mathematically
interesting option and I shall therefore in the rest of this Section
analyze some if its consequences, in particular for the behavior of
$F_2^{\mr{ep}}(x,Q)$ as a function of $Q$.

In the small $x$ region, to which we restrict our attention, the LO
expression for the distribution of sea quarks within a single quark
which at the scale $\mu$ is described by the initial distribution
$\delta(1-x)$, has the form \cite{DDT}
\be
D_0(x,Q)=\frac{1}{x}\frac{32}{3}C_2\zeta^2
\mr{e}^{-a\zeta}\frac{I_2(v)}{v^2},
\label{Dsea}
\ee
where for 3 colors and $n_f$ flavors
\be
a\equiv 11+\frac{2n_f}{27},\;\;
\zeta\equiv \frac{1}{2b}\ln\frac{a(\mu)}{a(Q)},\;\;
v\equiv\sqrt{48\zeta\ln(1/x)},\;\;\; C_2=\frac{4}{3}
\ee
and $I_2(v)$ is the modified Bessel function.
For general initial distribution $q(x,\mu)$ we have
\begin{eqnarray}
D(x,Q) & = & \int_x^1\frac{\mr{d}y}{y}q(y,\mu)D_0(x/y,Q)
\label{Dgeneral1} \nonumber \\
 & = & \frac{1}{x}\frac{32}{3}C_2\zeta^2\mr{e}^{-a\zeta}
\int_x^1\mr{d}yq(y,\mu)\frac{I_2(w)}{w^2} \label{Dgeneral2}
\end{eqnarray}
where now $w\equiv\sqrt{48\zeta\ln(y/x)}$.
Note that for fixed $x$ and $\zeta\rightarrow 0$, corresponding to
$Q\rightarrow \mu^{+}$, $v\rightarrow 0$ and $D_0(x,Q)$ behaves as
\be
D_0(x,Q)=\frac{1}{x}\frac{4}{3}C_2\zeta^2,
\label{limit0}
\ee
vanishing for $\zeta=0$, i.e. at $Q=\mu$. However, the physically
relevant case of fixed $Q>\mu$ and $x\rightarrow 0$ corresponds to
slightly more complicated limit $v\rightarrow \infty$. Eq.
\reff{Dsea} then implies
\be
D_0(x,Q^2)=\frac{1}{x}\frac{32}{3}C_2\zeta^2\mr{e}^{-a\zeta+v(x)}
\frac{1}{\sqrt{2\pi v(x)}}\frac{1}{v^2(x)}
\label{limit1}
\ee
where the terms depending on $v(x)$ induce additional dependence on $x$.
This modifies slightly the behavior of $D_0(x,Q)$ at small $x$, but does
not change its nonintegrability due to the dominant $\frac{1}{x}$
factor.  Consequently, $D_0(x,Q)$, considered as a function of $Q$ does
not vanish uniformly in the whole interval $x\in(0,1)$ when
$Q\rightarrow \mu$! The same holds for $D(x,Q)$. This nonuniform
convergence means that for $Q$ arbitrarily close to the input $Q_0=\mu$,
there is always a region of $x$ close to $x=0$, where approximately
$D(x,Q)\propto 1/x$.  And it is this region which causes the divergence
of the integral
\be
\int_0^1 \mr{d}x D(x,Q)=\infty
\label{div}
\ee
for any $Q>\mu$. In the conventional approach with
singular, i.e. nonintegrable, initial distributions, this discontinuity
is absent. Using eq.  \reff{Dgeneral2} and assuming for small $x$ quark
initial distribution in the form $q(x,\mu)=Ax^{-\lambda},\;\lambda>1$,
we find that for small $x$ and $Q\rightarrow \mu^{+}$ the integral
\be
\int^1_x\mr{d}yq(y,Q)\frac{I_2(w)}{w^2}
\label{intq}
\ee
behaves differently than for $\lambda<1$. For $\lambda>1$ the integrand
of \reff{intq} is a nonintegrable function of $x$ in the interval
$(0,1)$, finite lower integration bound is therefore crucial and we get
for any $Q>\mu$
\be
\int^1_x \mr{d}yq(y,Q)\propto \frac{A}{\lambda
-1}x^{1-\lambda}\;\;\;\; \Longrightarrow
\;\;\;\; D(x,Q)\propto \frac{1}{x}\int^1_x \mr{d}q(y,\mu)\propto
Ax^{-\lambda}.
\label{int2}
\ee
The singular initial distribution overrides the radiation pattern
characterizing the emissions from individual quarks and the radiated
sea quark distribution $D(x,Q)$ is therefore of the same form as the
initial $q(x,\mu)$. This well--known feature of singular initial
distributions implies that for low $x$ the form of the $x$--dependence
is essentially independent of $Q$ and the initial scale
plays no exceptional role.

\section{Conventional partons in small $x$ region}
In this section results of the conventional analysis based on singular
initial distributions will be cast into a simple form suitable for the
comparison with GRV results.  In order to check
the very essence of the GRV approach only the results based on the
original initial conditions \reff{initial} with vanishing antiquark and
gluon initial distributions will be discussed.  Moreover, I shall
concentrate on the small $x$ region, where the approximations of the
previous section are expected to hold \cite{DDT}.  In the approximation
of neglecting the second term in \reff{GGG} $H(x,M)\equiv xG(x,M)$
satisfies the equation
\be
\frac{\mr{d}H(x,M)}{\mr{d}\ln M}=a(M)\int^1_x
\frac{\mr{d}z}{z} \left(zP^{(0)}_{GG}(z)\right)H(x/z,M),
\label{Gdiag}
\ee
where
\be
P^{(0)}_{GG}(z)\equiv 6\left(\left[\frac{z}{1-z}\right]_{+}
+\frac{1-z}{x}+z(1-z)+\left(\frac{33-2n_f}{36}-1\right)\delta(1-z)
\right).
\label{P00GG}
\ee
Assuming $H(x,M)$ in a singular factorizable form
\be
H(x,M)=x^{-\lambda}F(M,\lambda), \;\;\lambda>0
\label{factorizable}
\ee
and substituting \reff{factorizable} into \reff{Gdiag} we get
\be
\frac{\mr{d}F(M,\lambda)}{\mr{d}\ln M}=
a(M)F(M,\lambda)\left(C(\lambda)-\int^x_0\mr{d}z
z^{\lambda}P^{(0)}_{GG}(z)\right)
\label{dH}
\ee
where
\be
\gamma^{(0)}(\lambda)
\equiv \int^1_0\mr{d}z z^{\lambda}P^{(0)}_{GG}(z)=
6\frac{\lambda(\lambda+1)+(\lambda+2)(\lambda+3)}
{\lambda(\lambda+1) (\lambda+2)(\lambda+3)}-\frac{3}{2}
-6\lambda\sum_{k=0}^{\infty}\frac{1}{(2+k)(\lambda+2+k)}.
\label{C}
\ee
extends the definition of the LO anomalous dimension $\gamma^{(0)}(n)$
to noninteger positive values of $\lambda$. Note that
$\gamma^{(0)}(\lambda)$ is a decreasing function of its
argument and negative for $\lambda$ above $\lambda_0\approx 0.85$.
The second term in the brackets of \reff{dH} is proportional to
$x^{\lambda}$ and can therefore be neglected
in the small $x$ region. $F(M,\lambda)$ then satisfies the equation
\be
\frac{\mr{d}F(M,\lambda)}{\mr{d}\ln M}=
a(M)\gamma^{(0)}(\lambda)F(M,\lambda),
\label{dF}
\ee
which has a simple solution
\be
F(M,\lambda)=A\left(a(M)\right)^{-\gamma^{(0)}(\lambda)/b},
\label{FM}
\ee
where $A$ is an arbitrary overall normalization constant.

In the small $x$ region the comparison of the GRV and conventional
results can be made particularly transparent by
translating \reff{FM} into the corresponding expression for the proton
structure function $F_2^{\mr{ep}}(x,Q^2)$ by means of a LO Prytz's
relation \cite{Prytz}
\be
\frac{\mr{d}F_2^{\mr{ep}}(\frac{x}{2},Q)}{\mr{d}\ln Q}=
\kappa a(Q)H(x,Q),\;\;\kappa=\frac{20}{27}.
\label{Prytz}
\ee
This formula was shown to be a good approximation for $F_2$ in the low
$x$ region and should be sufficient for our purposes.  Alternatively, we
could solve the DGLAP evolution equations for the coupled quark singlet
and gluon distributions exactly and then approximate these solutions by
the power--like behavior, but the procedure based on the combination of
\reff{FM} and \reff{Prytz} is much simpler. Anticipating also
$F_2^{\mr{ep}}$ in the factorizable form
\be
F_2^{\mr{ep}}(x,Q)=x^{-\lambda}F_2^{\mr{ep}}(Q,\lambda),
\label{F2M}
\ee
and using \reff{Prytz} in combination with the explicit result
\reff{FM} for $F(M,\lambda)$ we get
\be
\frac{\mr{d}F_2^{\mr{ep}}(M,\lambda)}{\mr{d}\ln M}=A\kappa
\left(\frac{1}{2}\right)^{\lambda}\left(a(M)\right)^
{-\gamma^{(0)}(\lambda)/b+1},
\label{dF2M}
\ee
wherefrom
\be
F_2^{\mr{ep}}(M,\lambda)=
\frac{\kappa}{2\gamma^{(0)}(\lambda)}F(M,\lambda)=
\frac{A\kappa}{\gamma^{(0)}(\lambda)}\left(\frac{1}{2}\right)^
{\lambda}\left(a(M)\right)^{-\gamma^{(0)}(\lambda)/b}.
\label{F2Msolution}
\ee
The positivity of $F_2^{\mr{ep}}(M,\lambda)$ requires positive 
$C(\lambda)$, which
in turn implies $\lambda<\lambda_0$. The measured behavior
of $F_2^{\mr{ep}}$ in the small $x$ region is well within this limit.

\section{DGPD vs. conventional partons -- numerical comparison}
Can the DGPD be distinguished from the conventional parton
parameterizations at all? In order to identify reasonably
unambiguous signatures of DGPD I shall concentrate on its
``orthodox'' version.

In the first kind of comparisons GRV results for
$F_2^{\mr{ep}}(x,Q^2)$ were obtained via \reff{F2ep}
from $u$ and $d$ quark valence--like initial distributions
\reff{uv}-\reff{dv}
\footnote{No essential features of the following comparisons
depend on this particular choice of the initial distributions.}.
The valence distributions were evolved by means of
the standard DGLAP evolution equations using the method of Jacobi
polynomials \cite{Barker,my}. For the sea parts \reff{Dgeneral2} was
used. Three light quarks were taken into account in generating the sea.
In order to facilitate the absolute comparison between the two
approaches without reference to experimental data,
the constants $A$ and $\lambda$ in \reff{FM} were related in such a way
that the conventional results coincide with the GRV ones for
$x=10^{-3}$ and $Q^2=5$ GeV$^2$. As a result $A$ becomes a function of
$\lambda$. The choice of the normalization point is, of course, to a
large extent arbitrary, but none of the basic messages
of this section depends on it. In Fig. 2a results of the GRV
approach are plotted for 10 values of
$Q^2=0.35,\;0.4,\;0.6,\;1,\;2,\;5,\;10,\;20,\;50,\;100$ GeV$^2$,
i.e. starting very close to the initial $\mu^2=0.34$ GeV$^2$.
The thick solid curve corresponds to the initial $F_2^{\mr{ep}}(x,\mu)$.
In Fig. 2b the corresponding sea distributions, which in the low $x$
region are well approximated by the power--like behavior of the form
$x^{-\lambda}$, are plotted.  Figs. 2a,b nicely illustrate the
way $F_2^{\mr{ep}}(x,Q)$ approaches the input function
$F_2^{\mr{ep}}(x,\mu)$ as $Q\rightarrow \mu^{+}$. In particular the
nonuniform convergence, discussed in Section 5, is clearly visible. As
$Q\rightarrow \mu^{+}$ the region of increasing $F_2^{\mr{ep}}$ moves
steadily to the left, but regardless of how close $Q$ is to $\mu$, such
a region eventually appears, leading to the discontinuity  of the
integral over quark distributions at the initial scale $\mu$.

Fitting for each $Q^2$ the total GRV sea contribution in
the interval $x\in(10^{-4},10^{-2})$ to the form $A(Q)x^{-\lambda(Q)}$
leads to characteristic dependence of the parameters $A(Q)$ and
$\lambda(Q)$ on $Q$, displayed in Figs. 3a,b by thick solid curves. The
corresponding power--like fits are shown as dotted curves in Figs. 2a,b.
As $Q\rightarrow\mu^{+}$ the interval where they provide good
approximation to full GRV results shifts systematically to smaller
values of $x$, reflecting the shift to the left of the relative
importance of the sea component.

In Fig. 4 GRV results are compared with those of the conventional
approach, shown as dashed lines, for six values of $\lambda$ and the
same values of $Q^2$ as in Fig. 2. The lowest, thick solid curves of the
GRV approach correspond again to the initial distributions 
\reff{uv}-\reff{dv} at 
$\mu^2=0.34$ GeV$^2$ and have thus no analog among the
dashed curves. Note that for smaller values of $\lambda$ not all curves
of the conventional approach fit into the frame of the plots. To
summarize the message of the Fig. 4, $F_2^{\mr{ep}}(Q,\lambda)$, defined
in \reff{F2M}, is plotted in Fig. 3a as a function of $Q^2$ for six
values of $\lambda$ together with the corresponding function $A(Q)$ of
the GRV approach.

In order to investigate the sensitivity of the parameters $\lambda(Q)$ 
and $A(Q)$ in the GRV approach to the choice of the initial 
distributions, I have repeated the calculations presented in Fig. 2a for 
the set of $u$ and $d$ quark initial distributions of the form
\be
xq_v(x,\mu)=Ax^{\alpha}(1-x)^{\beta}
\label{qform}
\ee
for two values of $\beta=3,4$ and five values of
$\alpha=0.1,0.3,0.5,0.7,0.9$. In all cases the overall normalization
factor $A$ was fixed to give two $u$ and one $d$ quark at the initial
scale $\mu$. Results corresponding to $\beta=3$ and the five mentioned
values of $\alpha$ are displayed in Fig. 5a-e. The general pattern
remains the same as in Fig. 2a, but there is a marked difference between
the dependence of the exponent $\lambda(Q)$ (curves included in Fig. 3b)
and the normalization factor $A(Q)$ (see Fig. 5f) on $\alpha$ and
$\beta$ (not shown). While the exponent $\lambda(Q)$ depends on the
values of both $\alpha,\beta$ very weakly, $A(Q)$ depends on $\alpha$
strongly and $\beta$ moderately.  The insensitivity of the exponent
$\lambda(Q)$ to details of the initial distributions can be understood
by closer inspection of the formula \reff{Dgeneral2}.
The figures 3--5 suggest two distinct features of the GRV results:
\bit
\item
The characteristic $Q^2$ dependence of the exponent $\lambda(Q)$.
In the conventional approach $\lambda$ is arbitrary
$Q^2$--independent number, while in GRV one it is an almost unique
function of $Q$, which starts rapidly from zero at the initial scale
$\mu$, but then progressively slows down. The most sensitive region is
clearly that close to the initial $\mu$, but even at large $Q^2$
the characteristic $Q^2$ dependence of GRV results persists. In the
region between $Q^2\doteq 2$ and $Q^2\doteq 100$ GeV$^2$, where leading
twist can perhaps be trusted and data from HERA are available,
$\lambda(Q^2)$ varies slowly in the range $(0.25,0.38)$. Unfortunately
the current experimental error on $\lambda$ at HERA is too large for a
reliable discrimination of this dependence from a constant one,
characterizing the conventional approach. To pin down the $Q^2$
dependence expected in the GRV approach reliably would require
lowering the experimental error on $\lambda(Q)$ to about $0.03$ in
a broad range of $Q^2$.

\item
The $Q^2$ dependence of the normalization factors $A(Q)$,
$F_2^{\mr{ep}}(Q,\lambda)$ respectively. Here the differences are much
more pronounced and increase as we go to both small and large values of
$Q^2$. They depend also much more on the choice of the initial
distributions. In GRV approach we cannot go below the initial scale,
while in the conventional approach there is no such strict limitation in
the small $x$ region.  Fig. 3a indicates that the region close to the
initial scale is particularly sensitive to the differences between the
two approaches.
\eit

The above comparisons show that the ``orthodox'' version of GRV
partons could be easily distinguished from the conventional
parameterizations, based on the singular initial distributions. However,
once valence--like initial gluons and antiquarks are added the
difference in the behavior of normalization factors $A(Q)$ and
$F_2^{\mr{ep}}(Q,\lambda)$ largely disappears. This is documented by the
fact that the latest GRV as well as the conventional parameterizations
can accommodate new HERA data \cite{H1,ZEUS} at low $x$ and in a broad
$Q^2$ range equally well. On the other hand the characteristic
$Q^2$ dependence of the exponent $\lambda(Q)$ remains essentially
unchanged as it is still given by the basic radiation pattern
\reff{Dsea}. Similarly even in the refined versions of DGPD
\cite{GRV4,GRV5} $F_2^{\mr{ep}}(x,Q)$ approaches the initial structure
function $F_2^{\mr{ep}}(x,\mu)$ in same nonuniform manner as discussed
in Section 5 and displayed in Figures 2 and 5.
The available experimental data either reach sufficiently low values
of $x\approx 10^{-4}$ but stop at $Q^2\ge 2$ GeV$^2$ (H1 and ZEUS at
HERA), or reach low $Q^2$ but only touch the crucial region of $x\le
10^{-3}$ (E665 at Fermilab). Despite its limited $x$ range recent E665
data indicate deviation from the latest GRV parameterizations for $x$
below $10^{-2}$ and $Q^2\le 1$ GeV$^2$ \cite{E665}. New quantitative
tests are expected from upgrades of H1 and ZEUS detectors, which will
extend the region of accessible $Q^2$ down to a fraction of GeV$^2$
and could soon throw some light on the questions discussed in this
paper.

\section{Summary and conclusions}
In this paper I have discussed the basic idea and consequences of
the DGPD from physical as well as mathematical points of view.
I have argued that it runs into problems if we attempt to interpret the
properties of initial distributions in physical terms. The only way to
avoid these problems, and the one adopted by GRV, is to assume
that the initial scale $\mu$ lies outside the range of validity of
leading twist perturbative QCD. All the peculiar properties discussed in
Sections 4,5 are then of no concern, but at the same time the approach
looses its physical justification and becomes primarily an exercise in
mathematics. Investigating the consequences of different initial parton
distributions on the solutions of LO/NLO DGLAP evolution equations is an
interesting mathematical problem of its own and I have therefore devoted
the second part of this paper to it. In particular I have looked
for signatures in the behavior of $F_2^{\mr{ep}}(x,Q^2)$ that would
provide clears signals that GRV dynamics is at work. The only, but
on the other hand rather unique signature of this kind is the
characteristic $Q^2$ dependence of the exponent $\lambda(Q)$ in
\reff{F2M}. To best way of pinning down this dependence at HERA is to
extend the $Q^2$ range below 1 GeV$^2$, where the variation of
$\lambda(Q)$ is strongest, and study in detail the $Q^2$ variation of
$\lambda(Q)$ in the whole available $Q^2$ range. But even then a
significant increase in the precision of measuring $\lambda$ is
necessary before a definite conclusion on the $Q^2$--dependence of
$\lambda$ can be drawn.

\newpage
\parindent 0.2cm
{\bf Figure captions}
\bd
\item {\bf Fig. 1:} The results of nonperturbative evaluation of
particularly defined QCD
running coupling $\alpha_s^{\mr{latt}}(\mu)$ on the lattice. Taken
from Ref. [19].
\item {\bf Fig. 2:} a) $F_2^{\mr{ep}}(x,Q^2)$ as a function of $x$ for
the initial distribution \reff{uv}-\reff{dv} and
10 values of $Q^2$ defined in the text. In the low $x$ region the
curves are ordered from below in order of increasing $Q^2$. The dotted
lines correspond to the power--like fits described in the text
and the thick solid curve describes the initial $F_2^{\mr{ep}}(x,\mu)$.
b) The same as in a), but for the total sea component only.
\item {\bf
Fig. 3:} $Q^2$--dependence of the normalization factor $A(Q)$ (a) and
the exponent $\lambda(Q)$ (b). In a) the thick solid curve is given by
the power--like fit of the GRV results, while the dashed lines,
describing $F_2^{\mr{ep}}(Q,\lambda)$ of the conventional approach,
correspond to six values of
$\lambda=0.05,\;0.1,\;0.2,\;0.3,\;0.4$ and $0.5$.
In b) the thick solid curve describes again the GRV result for the
initial distributions \reff{uv}-\reff{dv}, while the other curves
correspond (from above in decreasing order of $\alpha$) to five initial
distributions \reff{qform} with $\beta=3$ and
$\alpha=0.1,\;0.3,\;0.5,\;0.7,\;0.9$. In b) similar curves for
$\beta=4$ would be essentially indistinguishable from those in the
figure.
\item {\bf Fig. 4:}
The comparison of the GRV (solid curves) and conventional
(dashed curves) results for $F_2^{\mr{ep}}(x,Q^2)$ separately for
six values of $\lambda$ in the latter approach.
\item {\bf Fig. 5:} The same as in Fig. 2a (a--e) and Fig. 3a (f), but
for $\beta=3$ and five values of $\alpha$ in \reff{qform}.
In f) the curves are ordered from above in order of increasing $\alpha$.
\ed

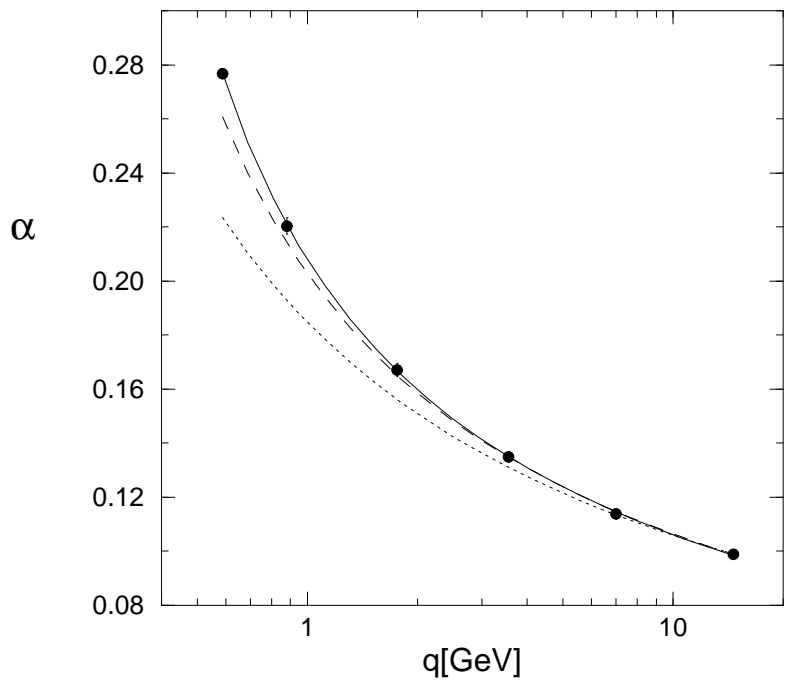
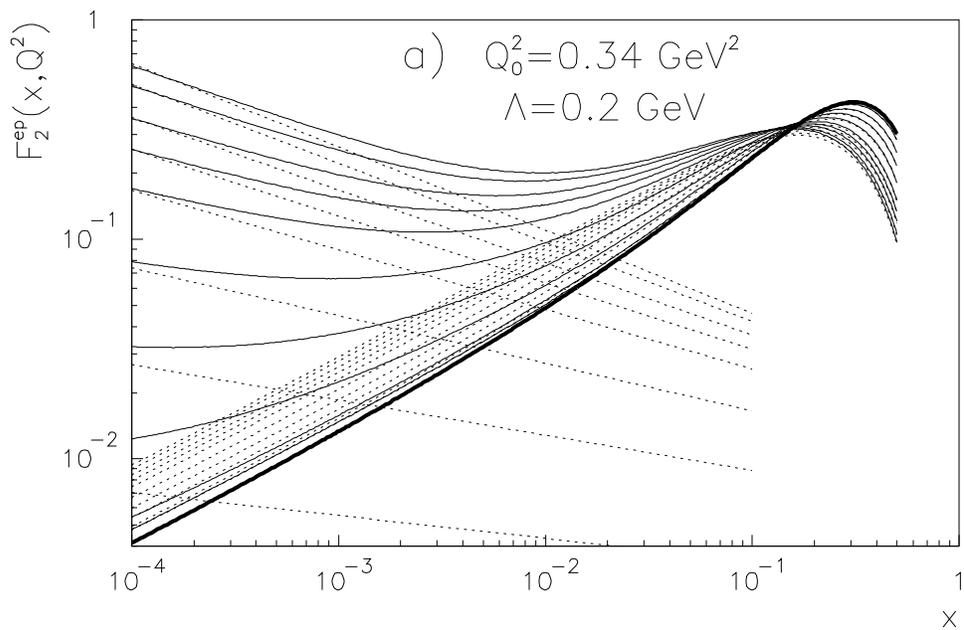
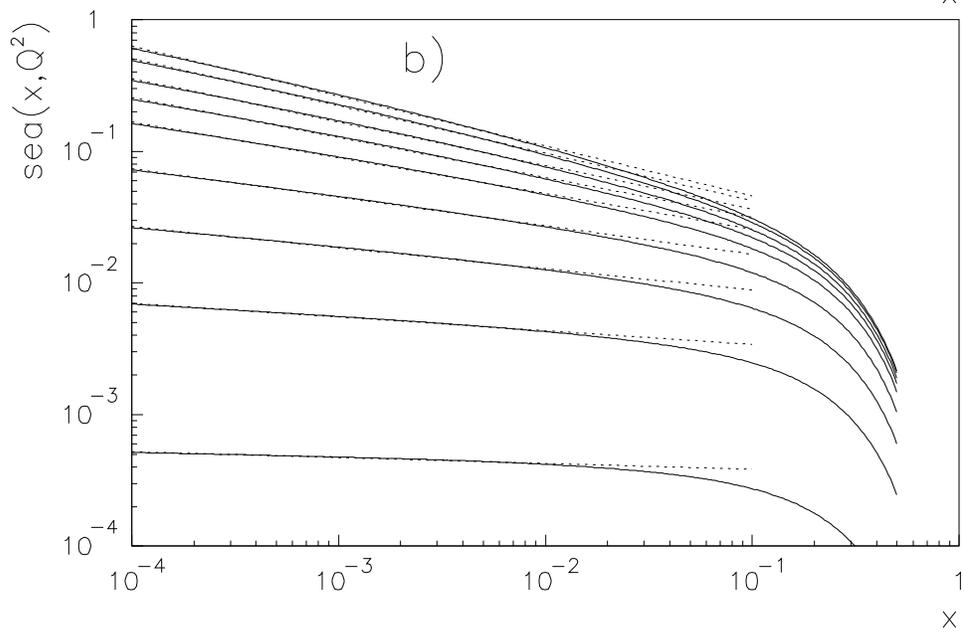
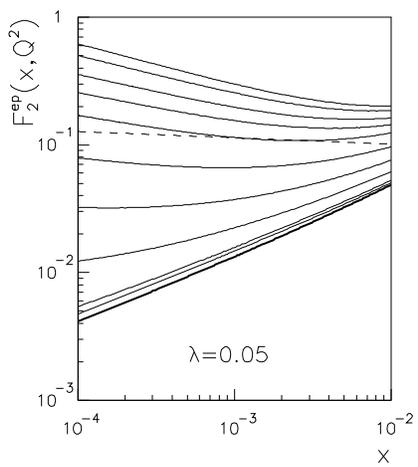
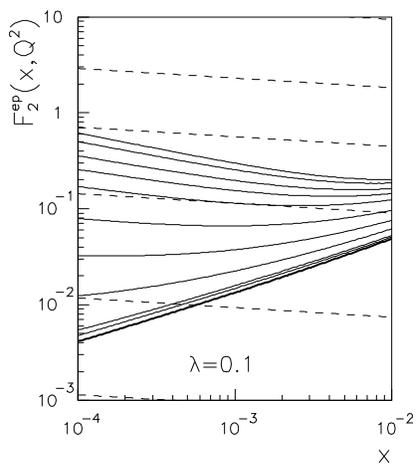
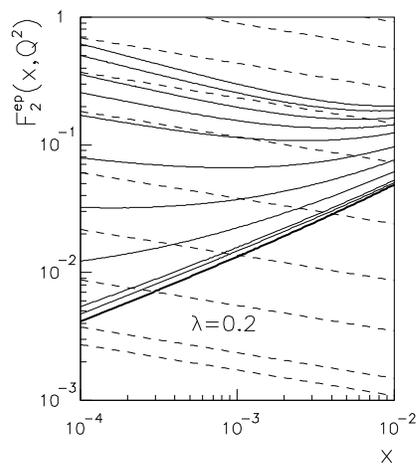
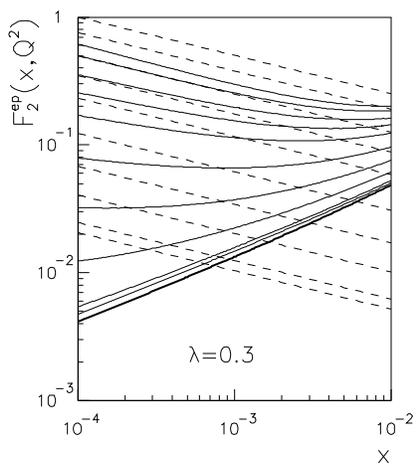
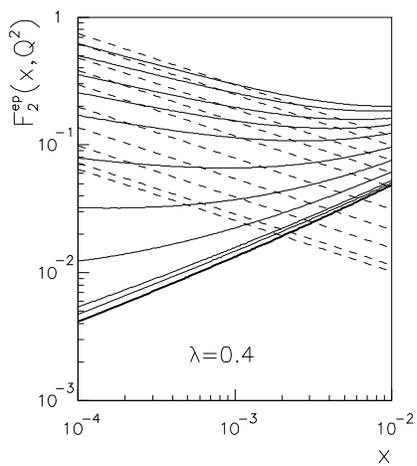
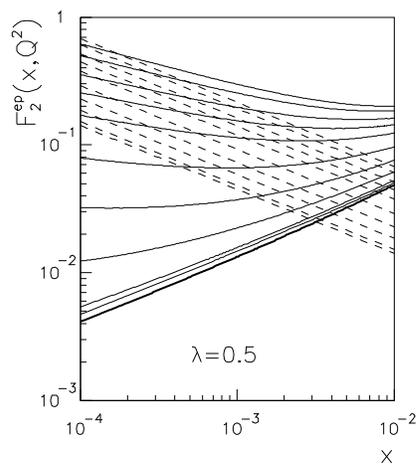
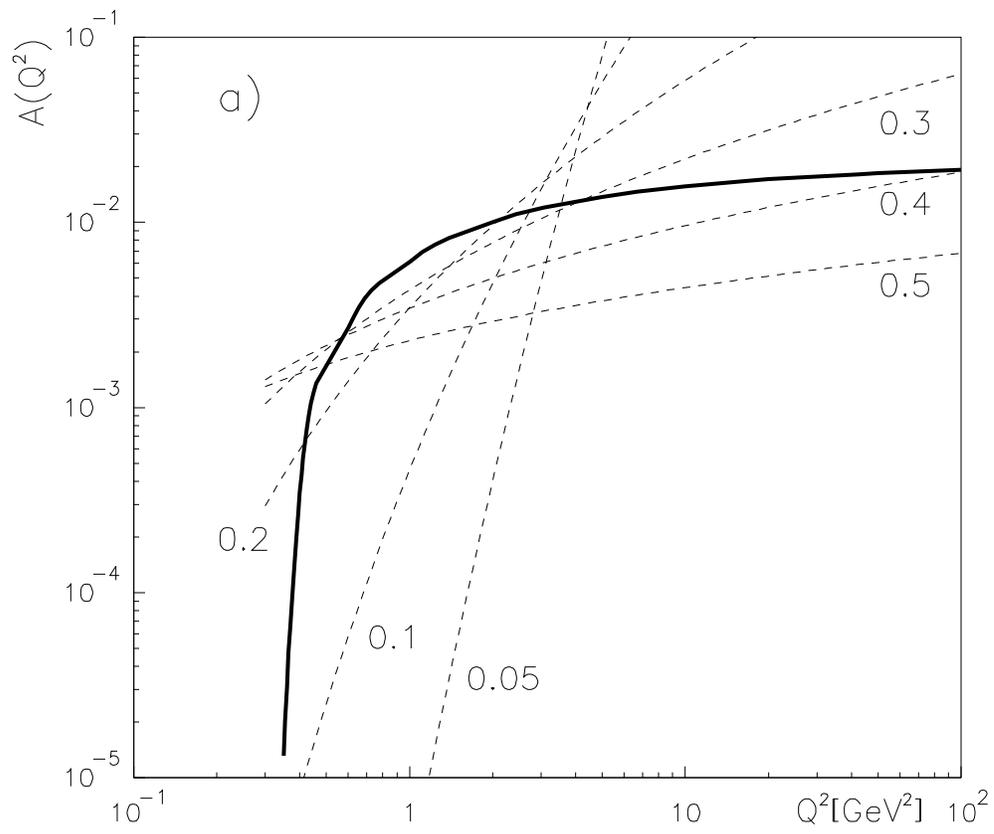
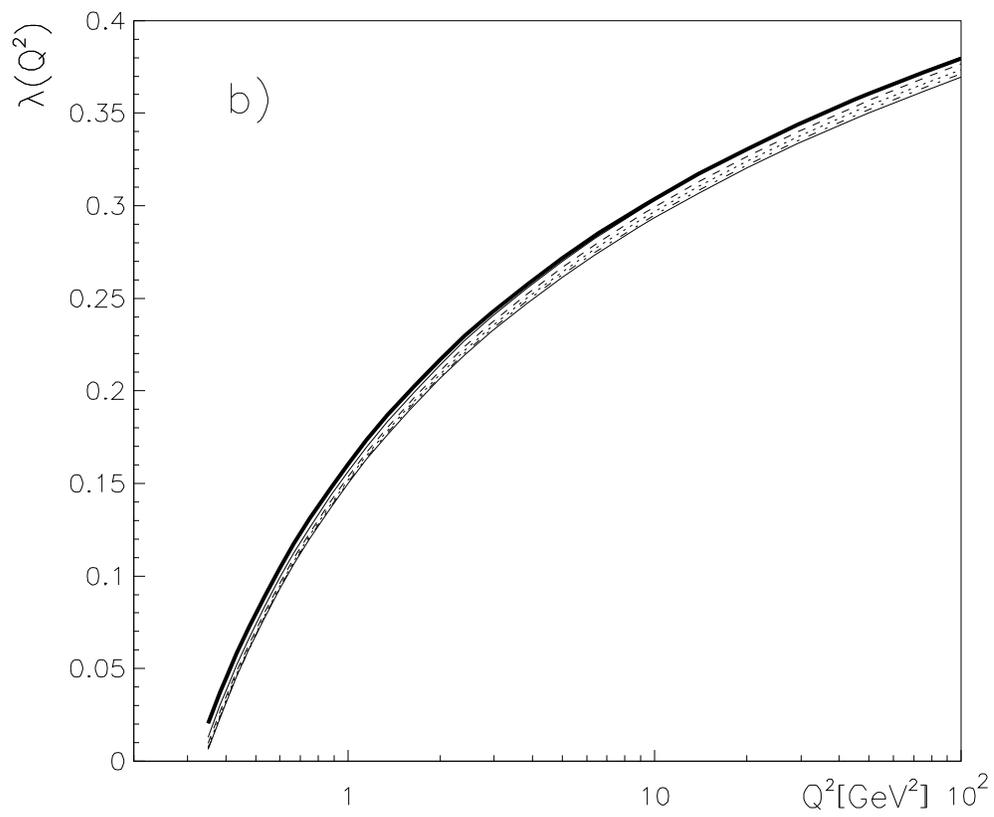
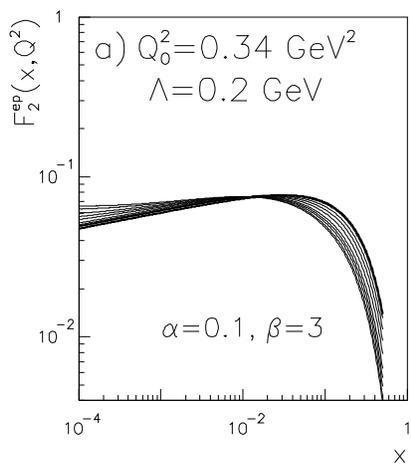
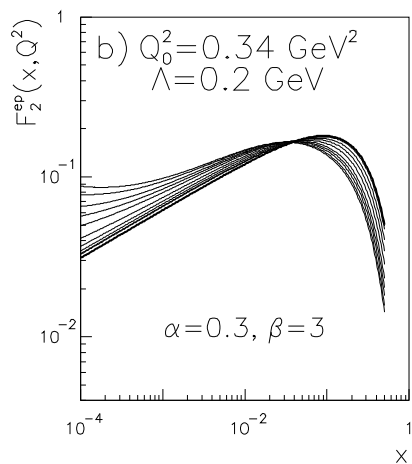
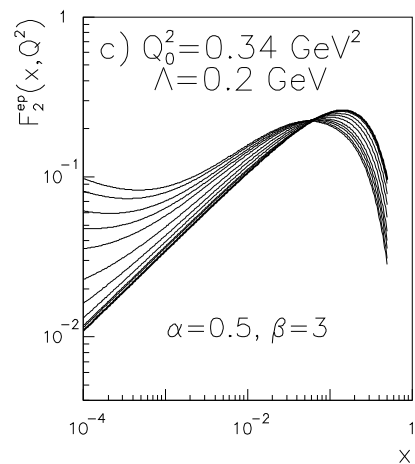
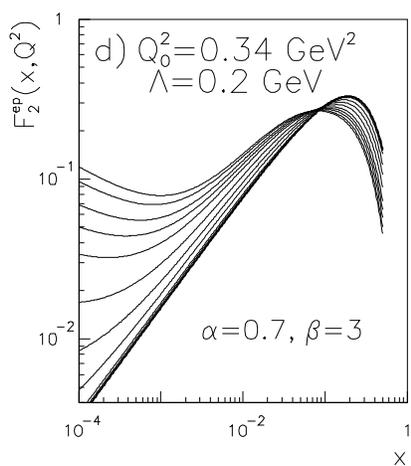
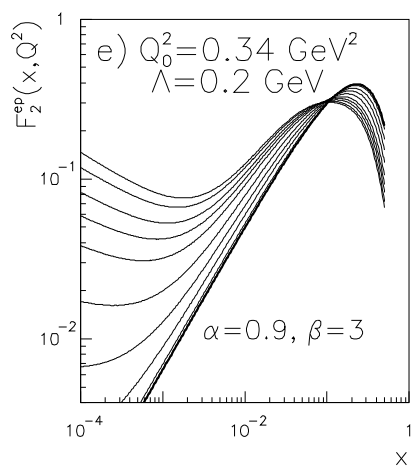
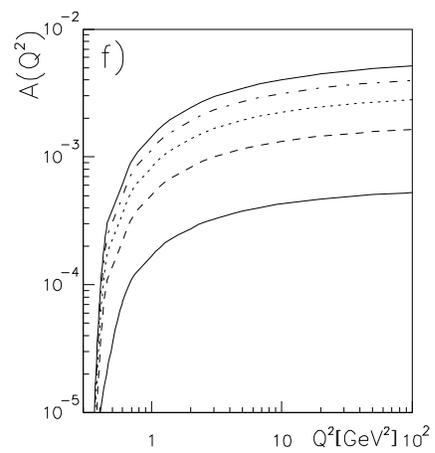
\end{document}